\documentclass[amsmath, amsfonts, showpacs, twocolumn, prl]{revtex4}
\usepackage{graphicx}
\usepackage{epsfig}
\usepackage{bm}
\usepackage{dcolumn}
\usepackage{amsmath}
\usepackage{amssymb}
\usepackage[varg]{txfonts}

\begin{document}

\title{Plasmon decay and thermal transport from spin-charge coupling in generic Luttinger liquids}

\author{Alex Levchenko}
\affiliation{Department of Physics and Astronomy, Michigan State
University, East Lansing, Michigan 48824, USA}

\begin{abstract}
We discuss the violation of spin-charge separation in generic nonlinear Luttinger liquids and investigate its effect on the relaxation and thermal transport of genuine spin-$1/2$ electron liquids in ballistic quantum wires. We identify basic scattering processes compatible with the symmetry of the problem and conservation laws that lead to the decay of plasmons into the spin modes. We derive a closed set of coupled kinetic equations for the spin-charge excitations and solve the problem of thermal conductance of interacting electrons for an arbitrary relation between the quantum wire length and spin-charge thermalization length.  
\end{abstract}

\date{December 20, 2014}


\maketitle

\textit{Introduction}.-- The most profound implication of the Luttinger liquid theory~\cite{Tomonaga,Luttinger,Haldane} is the separation between charge and spin degrees of freedom~\cite{Mattis,DL,Luther-Emery,Solyom}. The latter represent elementary low-energy excitations of the interacting spin-$1/2$ fermions which are bosonic waves of spin and charge densities. These collective modes do not interact and propagate without dispersion, i.e. with different velocities independent of the wave vector. The existence of the spin and charge branches in the excitation spectrum of a one-dimensional genuine electron liquid has been confirmed in momentum-resolved tunneling experiments in quantum wires~\cite{Auslaender-S05,Jompol-S09}. The effect is deduced from the electron tunneling probability spectra that exhibit sharp peaks at energies associated with the excitation of the two bosonic modes, which is viewed as a hallmark of spin-charge separation in the Luttinger liquids. There is also a growing interest in revealing effects associated with spin-charge separation in experiments with cold Fermi gases, where spin and charge refer to two internal atomic states and the atomic mass density, respectively~\cite{Recati-PRL03,Kollath-PRL05,Polini-PRL07,1D-FermiGas-Review}. 

The concept of spin-charge separation follows from an approximation made within the Luttinger liquid model, 
which assumes a strictly linear dispersion relation for electrons. In the generic case, spectrum curvature leads to a coupling between spin and charge modes. It is thus of special interest to investigate emergent phenomena that are beyond the scope of the conventional paradigm~\cite{Review}. 

A real time analysis of the evolution of single particle excitations and density wave packets,
based on the time-dependent density-matrix renormalization group method, revealed
the robustness of spin-charge separation beyond the low-energy limit of the conventional Luttinger liquid theory~\cite{Kollath-PRL05}. The ultimate fate of spin-charge separation has been recently addressed in studies of a spectral function and dynamical structure factor of spinful one-dimensional (1D) electrons~\cite{Teber-PRB07,Lante-Parola-PRB09,Schmidt-PRL-PRB10,Pereira-Sela-PRB10}. These studies showed that interaction-induced spin-charge separation survives away from the Fermi points in the sense that spectral functions still exhibit power-law threshold singularities, and that their behavior retains a certain universality. The power exponents of these singularities, however, differ from those in the conventional linear Luttinger liquid model. Furthermore, in the general case there appear qualitatively distinct features, for example, the charge-density structure factor acquires a peak at energies characteristic for the spin excitations.    
     
The consequences of spin-charge coupling at zero magnetic field for the transport properties of one-dimensional electron liquids have not been systematically studied. It should be expected that interaction between spin and charge degrees of freedom can lead to pronounced effects. Perhaps the most dramatic one is that neutral spin modes can mediate charge current~\cite{Brazovskii-JETPLett93,Nayak-PRB01}. Spin-charge coupling gives rise to a temperature-dependent contribution of the spin subsystem to the resistance of a quantum wire. At low temperatures, below the spin exchange constant, this contribution is exponentially small and the conductance of the wire remains quantized at $2e^2/h$. However at higher temperatures the conductance saturates to a new universal value $e^2/h$ since spin excitations are backscattered in the wire~\cite{Matveev}. Spin-charge coupling has also a sizable effect on the Coulomb drag resistivity between quantum wires, which cannot be described by the conventional Luttinger liquid model since it depends on violating particle-hole symmetry~\cite{Pereira-Sela-PRB10}. Large violation of the Wiedemann-Franz law in weakly disordered Luttinger liquids~\cite{FHK} becomes even more prominent due to the spin-charge coupling effect~\cite{Rosch-PRL09}. 

\textit{Motivation}.-- Another distinct feature of the Luttinger liquid model is the absence of inelastic scattering processes responsible for the relaxation of nonequilibrium states. At the level of the bosonic description of the model, this property is clear since the excitation spectrum for both spin and charge sectors can be written in terms of the quantum harmonic oscillator modes. In the original fermionic language this property is less obvious, since the Hamiltonian contains a four-fermion interaction term, but it follows from the constraints of momentum and energy conservations on the pair-particle collisions, which do not change the electron distribution function and thus do not cause relaxation. The leading effect stems from the consideration of three-particle collisions~\cite{Teber-EPJB06,Lunde-PRB07,Khodas-PRB07,Pereira-PRB09,Micklitz-PRB10,Karzig-PRL10,AL-PRB11a,AL-PRB11b,AL-PRL11,DGP-PRB12,Lamacraft-PRA13,Protopopov-arXiv14}.  
At weak interactions the three-particle scattering rate can be calculated by using the generalized Fermi golden rule in the $T$-matrix, and iterating the bare two-body interaction term to the second order. Such a perturbative approach assumes the Born condition for scattering, namely, that the typical excitation energy of particles should exceed the energy scale of interaction. In the spinful electron liquid at finite temperature this criterion is equivalent to the condition that temperature must exceed the energy scale set by the spin-charge separation. It is the subject of this paper to study relaxation and thermal transport in the opposite regime of the lowest temperatures where the picture of weakly interacting particles no longer applies and one has to develop the appropriate phenomenology to account for the effects of spin-charge coupling in the generic nonlinear Luttinger liquids. We elucidate the microscopic mechanism of relaxation due to plasmon decay and elaborate on its manifestation in the physical observables by computing the thermal conductance of electron liquids in ballistic quantum wires.  Apart from the conceptual significance, this study is also relevant for experiments. It has been recently observed that relaxation rates are parametrically distinct for particles and holes~\cite{Barak-NP10}. From the transport measurements it was reported that the value of the thermal conductance measured at the plateau of electrical conductance is smaller than the expected quantized value~\cite{Chiatti-PRL06} and that spin-charge separation leads to a strong violation of the Wiedemann-Franz law~\cite{Wakeham-NatCom11}. 

\begin{figure}
 \includegraphics[width=8cm]{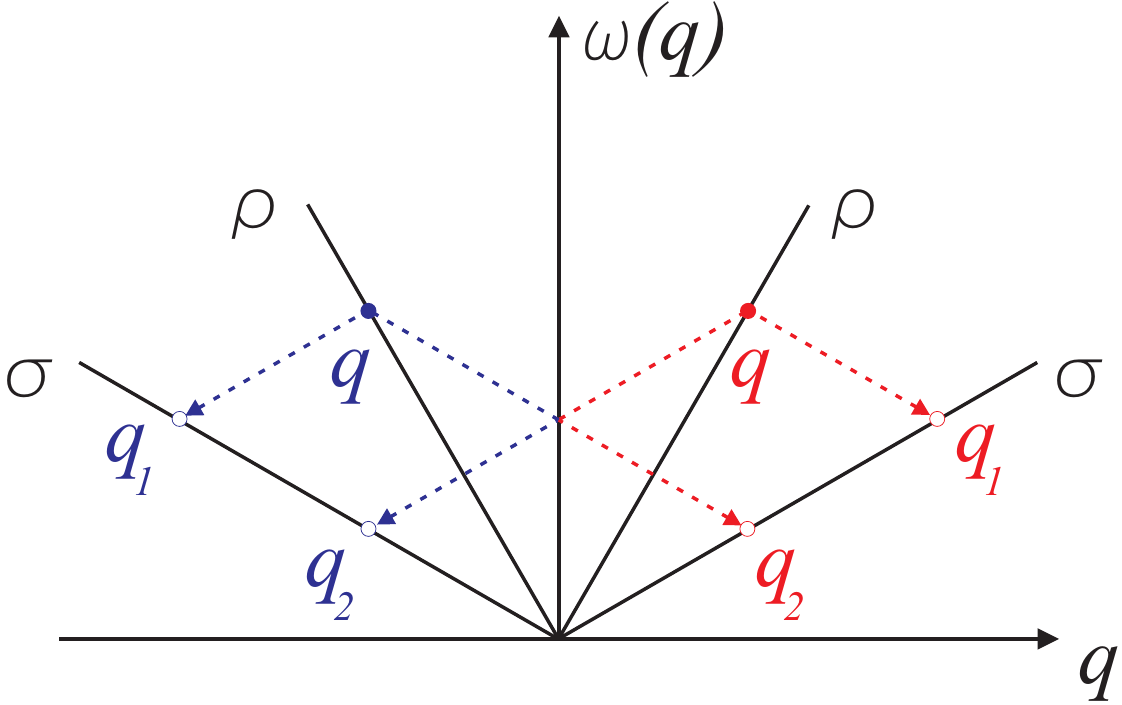}\\
 \caption{Schematic picture of an allowed decay of a plasmon with momentum $q$ into the counterpropagating spin excitations with momenta $q_{1,2}$. For the repulsive interaction $v_\rho>v_\sigma$ and kinematic constrains uniquely fix outgoing momenta: $q_1=q(v_\rho+v_\sigma)/2v_\sigma$ and $q_2=-q(v_\rho-v_\sigma)/2v_\sigma$. This scattering process emerges from the cubic nonlinearity of the spin-charge coupled Hamiltonian, Eq.~\eqref{H-3}.}\label{Fig1}
\end{figure}

\textit{Model and bosonization}.-- We consider interacting spin-$1/2$ fermions of mass $m$ in 1D with a quadratic dispersion relation described by the Hamiltonian (hereafter $\hbar=k_B=1$)
\begin{eqnarray}\label{H}
H=-iv_F\sum_s\int dx\left[\psi^\dag_{Rs}(x)\partial_x\psi_{Rs}(x)-\psi^\dag_{Ls}(x)\partial_x\psi_{Ls}(x)\right]\nonumber\\
-\frac{1}{2m}\sum_s\int dx\left[\psi^\dag_{Rs}(x)\partial^2_x\psi_{Rs}(x)+\psi^\dag_{Ls}(x)\partial^2_x\psi_{Ls}(x)\right]\nonumber\\
+\frac{1}{2}\sum_{ss'}\int dxdx' V(x-x')\psi^\dag_s(x)\psi^\dag_{s'}(x')\psi_{s'}(x')\psi_s(x).
\end{eqnarray}
Here the summation goes over the spin projection $s=\uparrow\downarrow$, $v_F$ is the Fermi velocity, and $V$ is the interaction potential. The annihilation field operators $\psi_{Rs}$ and $\psi_{Ls}$ represent right- and left-moving spin-$s$ electrons, while the full operator is $\psi_s=\psi_{Rs}+\psi_{Ls}$. We follow the usual prescription~\cite{Gogolin,Giamarchi} to bosonize this Hamiltonian by introducing $\psi_s(x)=e^{ik_Fx}R_s(x)+e^{-ik_Fx}L_s(x)$, where the new fields $R_s(x)$ and $L_s(x)$ are assumed to vary slowly on the scale of the Fermi wavelength $k^{-1}_F$.   
In the bosonization description these fields can be expressed in terms of bosonic displacement
$\varphi_s(x)$ and conjugated phase $\vartheta_s(x)$ as $R_s(x)=\frac{\kappa_s}{\sqrt{2\pi a}}e^{i\vartheta_s(x)-i\varphi_s(x)}$ and $L_s(x)=\frac{\kappa_s}{\sqrt{2\pi a}}e^{i\vartheta_s(x)+i\varphi_s(x)}$, where $a$ is the short distance cutoff $\sim k^{-1}_{F}$, and $\kappa_s$ are the Klein factors that ensure proper
anticommutation relations between original fermionic operators. They obey $\{\kappa_s,\kappa_{s'}\}=2\delta_{ss'}$ and satisfy $\kappa^\dag_s=\kappa_s$. The bosonic fields obey commutation $[\varphi_s(x),\vartheta_{s'}(x')]=\frac{i\pi}{2}\mathrm{sign}(x-x')\delta_{ss'}$. By transforming to the spin-charge representation $\varphi_\rho=\frac{1}{\sqrt{2}}(\varphi_\uparrow+\varphi_\downarrow)$, $\varphi_\sigma=\frac{1}{\sqrt{2}}(\varphi_\uparrow-\varphi_\downarrow)$, and similarly for the conjugated $\vartheta$ field, being careful with the point splitting of the operators, and keeping track of the leading order nonlinearities, we arrive at the following Hamiltonian   
$H=H_\rho+H_\sigma+H_{3}+H_4$~\cite{SM}:
\begin{equation}\label{H-rho}
H_\rho=\frac{v_\rho}{2\pi}\int dx\left[K^{-1}_\rho(\partial_x\varphi_\rho)^2+K_\rho(\partial_x\vartheta_\rho)^2\right],
\end{equation}
\begin{eqnarray}\label{H-sigma}
H_\sigma=\frac{v_\sigma}{2\pi}\int dx\left[K^{-1}_\sigma(\partial_x\varphi_\sigma)^2+K_\sigma(\partial_x\vartheta_\sigma)^2\right]\nonumber\\+\frac{g_\sigma}{2\pi^2a^2}\int dx\cos\big[2\sqrt{2}\varphi_\sigma\big],
\end{eqnarray}
\begin{eqnarray}\label{H-3}
H_{3}=\!\!\int\!\! dx\left[\eta(\partial_x\varphi_\rho)^3+3\eta(\partial_x\varphi_\rho)(\partial_x\varphi_\sigma)^2+
3\zeta(\partial_x\varphi_\rho)(\partial_x\vartheta_\rho)^2\right.\nonumber\\
\left.+3\zeta(\partial_x\varphi_\rho)(\partial_x\vartheta_\sigma)^2+6\zeta(\partial_x\vartheta_\rho)(\partial_x\varphi_\sigma)
(\partial_x\vartheta_\sigma)\right],
\end{eqnarray}
\begin{equation}\label{H-4}
H_4=\lambda\int dx \left[(\partial_x\varphi_\rho)^4+6(\partial_x\varphi_\rho)^2(\partial_x\varphi_\sigma)^2+(\partial_x\varphi_\sigma)^4\right].
\end{equation}
The conventional two terms $H_\rho$ and $H_\sigma$ describe the excitations of the charge and spin degrees of freedom, which are decoupled. The other two terms, $H_3$ and $H_4$, and also higher order operators, mix spin and charge modes, and thus capture interactions between bosons. These terms appear upon careful bosonization of the band curvature and backscattering  terms in the original Hamiltonian, Eq.~\eqref{H}. The parameters $K_{\rho(\sigma)}, g_\sigma, \eta, \zeta$, and $\lambda$ are determined by the interactions between electrons, while $v_{\rho(\sigma)}=v_F/K_{\rho(\sigma)}$ are the velocities of propagation of spin and charge excitations. For repulsive interactions $v_\rho>v_\sigma$. When deriving the bosonized Hamiltonian we assumed relatively weak interactions between electrons. One finds at that limit to the leading order in $V$:
$K_\rho=1-\frac{2V_0-V_{2k_F}}{2\pi v_F}$, $K_\sigma=1+\frac{V_{2k_F}}{2\pi v_F}$, $g_\sigma=V_{2k_F}$, $\eta=-\frac{1}{6\sqrt{2}\pi m}+\frac{V'_{2k_F}}{3\sqrt{2}\pi^2}$, $\zeta=-\frac{1}{6\sqrt{2}\pi m}$, and $\lambda=-\frac{V''_{2k_F}}{12\pi^2}$, where $V_0$ and $V_{2k_F}$ are zero-momentum and $2k_F$ Fourier components of the interaction potential. We note that the form of the Hamiltonian Eqs.~\eqref{H-rho}-\eqref{H-4}, is dictated by the SU(2) symmetry and thus is universal. It is thus expected to describe the low-energy properties of generic 1D electron liquids with arbitrarily strong interactions. The interaction parameters of the Hamiltonian can be fixed phenomenologically by relating them to the other observable quantities~\cite{Schmidt-PRL-PRB10,Pereira-Sela-PRB10}. For the SU(2) symmetric point the spin-flip coupling constant $g_\sigma$ scales to zero at low-energy scales $\varepsilon$ as $g_\sigma\to g_\sigma/[1+(g_\sigma/\pi v_\sigma)\ln(k_Fv_\sigma/\varepsilon)]$. The parameter $K_\sigma=1+g_\sigma/2\pi v_\sigma$ renormalizes along with it to unity $K_\sigma=1$.   

\textit{Kinetic equations}.-- We concentrate on the anharmonic terms in the Hamiltonian, Eqs.~\eqref{H-3}-\eqref{H-4}. An inspection of the kinematic constraints reveals that the cubic nonlinearity in boson fields allows a decay process of a plasmon into counterpropagating spin excitations Fig.~\ref{Fig1}. One should note that similar physics of a plasmon decay into neutral modes has been discussed in the context of carbon nanotubes~\cite{Wei-PRB10}. Quartic nonlinear terms in Eq.~\eqref{H-4} are of the same order in interaction and also lead to plasmon decay. However, they include four bosons in a scattering process and the corresponding rate is parametrically smaller in $q/k_F\ll1$ than that due to cubic nonlinearity because of phase space limitations. Spin excitations cannot decay unless curvature in their spectrum is accounted for explicitly.  

Our goal is to explore the consequences of spin-charge collisions on the kinetics of 1D electron liquids. For this purpose, we expand the bosonic fields in normal modes       
\begin{eqnarray}
\hskip-.25cm&&
\partial_x\varphi_\nu(x)=-\sqrt{\frac{\pi}{2\ell}}\sum_q\!\!\sqrt{|q|}e^{-iqx}\left[b^\dag_\nu(q)+b_\nu(-q)\right],\\
\hskip-.25cm&&\partial_x\vartheta_\nu(x)=\sqrt{\frac{\pi}{2\ell}}\sum_q\!\!\sqrt{|q|}\mathrm{sign}(q)e^{-iqx}\left[b^\dag_\nu(q)-b_\nu(-q)\right],\end{eqnarray}
for $\nu=\rho,\sigma$, where $\ell$ is the system size, and derive a closed set of coupled kinetic equations for the occupation functions $N_{\rho(\sigma)}(q)$ of spin and charge modes. For the process depicted in Fig.~\ref{Fig1}  we obtain for the stationary but spatially nonuniform situation
\begin{equation}\label{KE}
\pm v_\nu \partial_xN^{R(L)}_\nu=\mathrm{St}\{N_\nu\},
\end{equation}
where the plus/minus sign stands for the right/left movers, respectively, and the collision integral reads
\begin{equation}\label{St}
\mathrm{St}\{N_\rho\}=-W\left[N^R_\rho(1+N^R_{\sigma1})(1+N^L_{\sigma2})-(1+N^R_\rho)N^R_{\sigma1}N^L_{\sigma2}\right]\end{equation} 
with $\mathrm{St}\{N_\sigma\}=-\mathrm{St}\{N_\rho\}$. The notations here are such that $N_{\sigma1,2}=N_\sigma(q_{1,2})$, and the momenta of the outgoing spin waves are uniquely fixed by the momentum and energy conservations $q_{1}=q(v_\rho+v_\sigma)/2v_\sigma$ and $q_2=-q(v_\rho-v_\sigma)/2v_\sigma$. The scattering rate in Eq.~\eqref{St} that follows from Eq.~\eqref{H-3} is given by 
\begin{equation}\label{W}
W(q)=\frac{|q|^3K_\rho(V'_{2k_F})^2}{64v_\sigma}\frac{v^2_\rho-v^2_\sigma}{v^2_\sigma}.
\end{equation}
One should emphasize here that band curvature terms $\propto 1/m$ cancel out from the scattering rate \eqref{W}, which is thus governed solely by the interaction terms.  $W(q)$ can be associated with the plasmon attenuation coefficient (inverse life time $\tau^{-1}_\rho\propto W$). For the weakly interacting limit one can take $v_\rho-v_\sigma\sim V_0\ll v_F$ and estimate $V'_{2k_F}\sim V_{2k_F}/k_F$. This results in
$\tau^{-1}_\rho\sim E_F(V_0/v_F)(V_{2k_F}/v_F)^2(q/k_F)^3$~\cite{Schmidt-PRL-PRB10,Pereira-Sela-PRB10}, where $E_F$ is the Fermi energy. An interesting feature of this estimate is that the scattering rate scales as a third power of the interaction parameter, which is nonanalytic in the sense of perturbation theory in $V$. This peculiarity should be understood as the result of perturbation theory in the limit of weak backscattering constructed on the basis of the well-defined charge and spin modes, namely, when $V_{2k_F}\ll V_0\ll v_F$. It should also be remarked that nonlinearity of the plasmon dispersion relation results in two-plasmon collisions that give rise to a finite lifetime of charge modes even in the spinless case~\cite{Lin-PRL13,AL-PRB13}. The corresponding rate for this process is of higher order in $q/k_F\ll1$, namely, $\tau^{-1}_\rho\propto(q/k_F)^5$, so that the spin-charge coupling effect is expected to dominate the attenuation of plasmons. Dispersion nonlinearity of the spin excitations should also lead to attenuation of spin waves due to spin-spin collisions but this problem has not been addressed. Spin relaxation has been discussed in the context of a decay of spin currents due to backscattering spin-flip interaction which is present already in the linear LL model but assumes spin imbalance due to polarizing magnetic field or ferromagnetic leads~\cite{Balents-Egger-PRB01}. At weak interaction the corresponding relaxation rate is linear in temperature $\tau^{-1}_{\sigma}\propto(V_{2k_F}/v_F)^2T$ except for the lowest temperatures where this rate is suppressed exponentially. 
 
 \textit{Thermal conductance}.-- We apply the kinetic equations \eqref{KE}-\eqref{St} to calculate plasmon-assisted thermal transport in 1D electron liquids. Consider a quantum wire of length $\ell$ attached to leads that are kept at different temperatures $T_{l,r}$. To find the thermal conductance $\mathcal{K}$ of the system we need to solve the kinetic equations to linear order in the temperature difference $\Delta T=T_l-T_r$. To this end, we linearize Eq.~\eqref{KE} by parametrizing the distribution functions as follows 
 \begin{equation}\label{N-Phi}
 N^{R(L)}_\nu(q,x)=n_\nu+n_\nu(1+n_\nu)\Phi^{R(L)}_\nu(q,x)
 \end{equation} 
 where $n_\nu=[e^{\omega_\nu/T}-1]^{-1}$ is the equilibrium Bose distribution function with $\omega_\nu=v_\nu|q|$, while the perturbation is $\Phi_\nu\propto\Delta T$. This particular choice of $N^{R(L)}_\nu$ conveniently takes care of the detailed balance condition. We infer boundary conditions by assuming that the temperatures of the right- and left-moving plasmons near the ends of a wire $x=0,\ell$ are controlled by those in the leads, which implies 
 \begin{equation}
 \Phi^R_\nu(q,0)=\frac{\omega_\nu\Delta T}{2T^2},\quad \Phi^L_\nu(0,\ell)=-\frac{\omega_\nu\Delta T}{2T^2}. 
 \end{equation}  
 
First, we consider the decay of a right moving plasmon. This amounts to solving a set of three coupled linear differential equations for $N^R_\rho$, $N^R_{\sigma_1}$, and $N^L_{\sigma2}$. In the parametrization of Eq.~\eqref{N-Phi} we find 
\begin{equation}\label{Phi-r}
\Phi^R_\rho(x)=\frac{\omega_\rho\Delta T}{2T^2}+\frac{v_\sigma}{v_\rho}\frac{n_{\sigma2}(1+n_{\sigma2})}
{n_\rho(1+n_\rho)}[\Phi^L_{\sigma2}(x)-\Phi^L_{\sigma2}(0)],
\end{equation}   
\begin{equation}\label{Phi-s1}
\Phi^R_{\sigma1}(x)=\frac{\omega_{\sigma1}\Delta T}{2T^2}-\frac{n_{\sigma2}(1+n_{\sigma2})}
{n_{\sigma1}(1+n_{\sigma1})}[\Phi^L_{\sigma2}(x)-\Phi^L_{\sigma2}(0)],
\end{equation}    
\begin{eqnarray}\label{Phi-s2}
&&\hskip-.5cm
\Phi^L_{\sigma2}(x)=-\frac{\omega_{\sigma2}\Delta T}{2T^2}\frac{e^{-x/\xi}-\gamma}{e^{-\ell/\xi}-\gamma}\nonumber\\
&&\hskip-.5cm+\frac{(\omega_\rho-\omega_{\sigma1})\Delta T}{2T^2}\frac{e^{-\ell/\xi}-1}{1-\gamma}
\left[\frac{e^{-x/\xi}-\gamma}{e^{-\ell/\xi}-\gamma}-\frac{e^{-x/\xi}-1}{e^{-\ell/\xi}-1}\right].
\end{eqnarray}
In these equations we introduced 
\begin{eqnarray}
&&\gamma=\frac{n_{\sigma2}(1+n_{\sigma2})}{n_{\sigma1}(1+n_{\sigma1})}+\frac{v_\sigma}{v_\rho}\frac{n_{\sigma2}(1+n_{\sigma2})}{n_\rho(1+n_\rho)},\\
&&\xi^{-1}=\frac{(\gamma-1)W(1+n_\rho)n_{\sigma1}}{v_\sigma(1+n_{\sigma2})}.
\end{eqnarray}

Second, we consider decay of a left moving plasmon, which is not equivalent to the above considered case because of the finite thermal bias that manifestly breaks the detailed balance condition. The solution of the kinetic equations is analogous and can be obtained from Eqs.~\eqref{Phi-r}-\eqref{Phi-s2} by replacing $R\to L$, $\Delta T\to-\Delta T$, and $x\to \ell-x$. With the help of the distribution functions we define now the heat current 
\begin{equation}\label{Jq}
J_Q=\sum_{\nu=\rho,\sigma}\int^{\infty}_{0}\frac{dq}{2\pi} v_\nu \omega_\nu \left[N^R_\nu(q,x)-N^L_\nu(q,x)\right],
\end{equation}
which can be split into the two terms $J_Q=J_0+\delta J_Q$. The first one is just a current of noninteracting particles $J_0=\mathcal{K}_0\Delta T$ with thermal conductance $\mathcal{K}_0=2\pi^2 T/3$. All the spin-charge interaction effects can be absorbed into the second term, $\delta J_Q$. When computing $\delta J_Q$ one has to recalculate the nonequilibrium distributions $\Phi^{R(L)}_{\sigma}$ from their respective momenta $q_{1,2}$ to the running integration momentum $q$ in Eq.~\eqref{Jq}. Technically this requires inclusion of the whole series of scattering terms similar to that in Fig.~\ref{Fig1}. The final result can be expressed in terms of the distribution function of the lowest energy spin excitation $\delta J_Q=-\int\frac{dq}{2\pi}v_\sigma\omega_{\sigma2}n_{\sigma2}(1+n_{\sigma2})\big[\delta\Phi^L_{\sigma2}-\delta\Phi^R_{\sigma2}\big]$, where $\delta\Phi^{L}_{\sigma2}=\Phi^{L}_{\sigma2}(0)-\Phi^L_{\sigma2}(\ell)$ and $\delta\Phi^{R}_{\sigma2}=\Phi^{R}_{\sigma2}(\ell)-\Phi^R_{\sigma2}(0)$. By using Eqs.~\eqref{Phi-r}-\eqref{Phi-s2} $\delta J_Q$ can be reduced to
\begin{equation}
\delta J_Q=-\frac{2\Delta T}{T^2}\int^{\infty}_{0}\frac{dq}{2\pi}v_\sigma\omega^2_{\sigma2}n_{\sigma2}(1+n_{\sigma2})\frac{e^{-\ell/\xi}-1}{e^{-\ell/\xi}-\gamma}.
\end{equation}
It is an important check to see that the heat current is uniform along the wire and all coordinate dependent terms in the distribution functions cancel out.  

\begin{figure}
 \includegraphics[width=8cm]{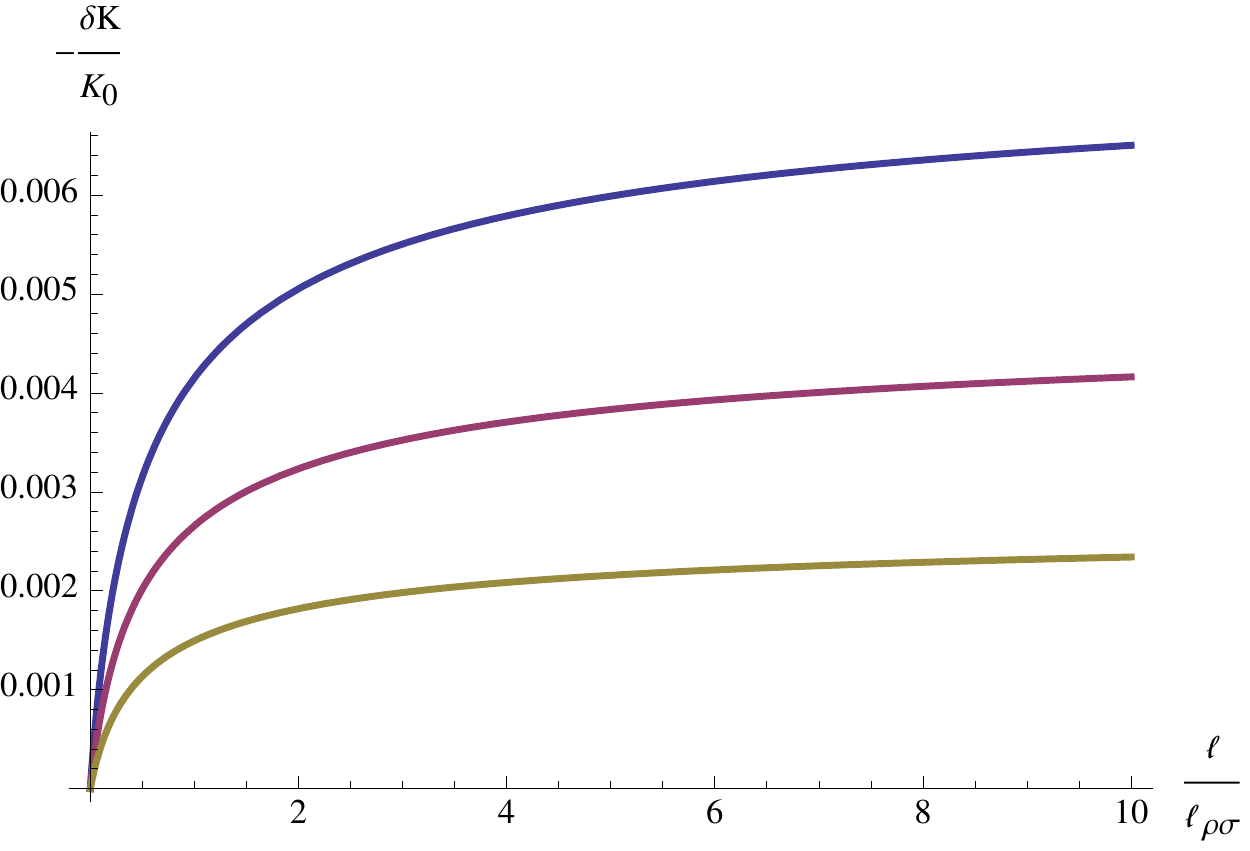}\\
 \caption{Interaction-induced correction to the thermal conductance of a clean quantum wire as a function of its length plotted for different values of interaction strength encoded by the difference between charge and spin velocities (from the bottom to the top curve) $(v_\rho-v_\sigma)/v_\sigma=0.15,0.2,0.25$. For $\ell\ll\ell_{\rho\sigma}$ the correction scales with $\ell$ and saturates to a constant value proportional to $(v_\rho-v_\sigma)^2/v^2_\sigma$ once $\ell\gg\ell_{\rho\sigma}$ in accordance with Eq.~\eqref{K}.}\label{Fig2}
\end{figure}

The typical momentum change of a plasmon in collisions is set by the temperature $q\sim T/v_\rho$. Provided the condition $v_\rho-v_\sigma\ll v_F$ one immediately concludes that $q\sim q_1\gg q_2$. This observation allows us to expand $n_{\sigma2}\approx T/\omega_{\sigma2}$, approximate $\xi^{-1}\approx q^2T(V'_{2k_F})^2/8v^3_\sigma$, and
neglect $e^{-\ell/\xi}$ compared to $\gamma$ in the denominator of Eq.~\eqref{Jq} since $\gamma(q\to0)\to8v^2_\sigma/(v_\rho-v_\sigma)^2\gg1$. This helps to simplify the expression for $\delta J_Q$ and find the correction to the thermal conductance $\delta\mathcal{K}=\delta J_Q/\Delta T$ in the form 
\begin{equation}\label{K}
\frac{\delta\mathcal{K}}{\mathcal{K}_0}=-\frac{3}{4\pi^2}\left(\frac{v_\rho-v_\sigma}{v_\sigma}\right)^2
\int^{\infty}_{0}\frac{z^2dz}{\sinh^2z}\left[1-e^{-z^2\ell/\ell_{\rho\sigma}}\right],
\end{equation}
where we have introduced the spin-charge thermalization length 
$\ell^{-1}_{\rho\sigma}=T^3(V'_{2k_F})^2/2v^5_\sigma$. For short wires $\ell\ll\ell_{\rho\sigma}$ the interaction-induced correction to the thermal conductance scales linearly with the wire length $\delta\mathcal{K}/\mathcal{K}_0=-(\pi^2\ell/40\ell_{\rho\sigma})(v_\rho-v_\sigma)^2/v^2_\sigma$. For long wires, $\ell\gg\ell_{\rho\sigma}$, it saturates to the temperature and length independent value $\delta\mathcal{K}/\mathcal{K}_0=-(v_\rho-v_\sigma)^2/8v^2_\sigma$. The correction term in $\delta\mathcal{K}$ at $\ell\gg\ell_{\rho\sigma}$ falls off algebraically, $\propto\sqrt{\ell_{\rho\sigma}/\ell}$. Figure~\ref{Fig2} represents the behavior of $\delta\mathcal{K}$ as a function of the wire length.   

It is instructive to compare these results to the earlier calculations of thermal conductance in 1D electron liquids at weak interaction from three-particle collisions~\cite{AL-PRB11b}. The present results apply to the regime of temperatures below the energy scale of spin-charge separation, $T<T_{\rho\sigma}\sim k_F(v_\rho-v_\sigma)$. 
Above that scale the saturated value of the thermal conductance crosses over to $\delta\mathcal{K}/\mathcal{K}_0\sim-(T/E_F)^2$ in agreement with Ref.~\cite{AL-PRB11b}. The thermalization length $\ell_{\rho\sigma}$ crosses over to $\ell^{-1}_{\rho\sigma}\sim(V_0/v_F)^2(V_{2k_F}/v_F)^2(T/v_F)$, also in agreement with the calculations of Ref.~\cite{AL-PRB11b}. To see that, one has to replace $T^2\to k^2_F(v_\rho-v_\sigma)^2\sim k^2_F V^2_0$  and estimate $V'_{2k_F}\sim V_{2k_F}/k_F$ in $\ell_{\rho\sigma}$. For the typical parameters of quantum wires used in the experiments~\cite{Barak-NP10} one estimates the thermalization length to be on the scale of a few micrometers~\cite{Note}.  

\textit{Summary}.-- We have studied emergent transport phenomena in generic Luttinger liquids based on the bosonized Hamiltonian of spin-1/2 electrons beyond the conventional limit. We discussed anharmonic perturbations associated with the band curvature and the backscattering effects that mix charge and spin excitations. Violation of the spin-charge separation, combined with kinematic constraints and SU(2) symmetry of the problem allow a decay process of plasmons into neutral spin modes. Spin waves can also decay but this is a higher order effect which requires consideration of nonlinearities in their dispersion relation. We conclude quite generally that relaxation processes in 1D electron liquids are hierarchical and characterized by the multiple time scales.  Attenuation of the plasmons leads to thermalization and modification of the thermal conductance of interacting electron liquids. Our main results are Eqs.~\eqref{St}, \eqref{W}, and \eqref{K}, which represent the collision integral due to spin-charge coupling, the scattering rate for plasmons and the thermal conductance, respectively.   

\textit{Acknowledgments}.-- I would like to thank K.~Matveev for the discussions that initiated this work and Z.~Ristivojevic for participation at the early stages of the project. I am also grateful to R.~Pereira and E.~Sela for the explanations regarding their work~\cite{Pereira-Sela-PRB10}, and to T.~Micklitz for discussion of the results. I am thankful to N.~Birge for reading and commenting on the manuscript. This work was supported by NSF Grant No. DMR-1401908.


\begin{thebibliography}{99}
\bibitem{Note}
For the numerical estimates, one uses typical parameters from the experiment of Ref.~\cite{Barak-NP10}. In particular, $v_F\sim2.5\times10^5$ m/s, $k_F\sim10^8$ m$^{-1}$, $\kappa\sim10$ is the dielectric constant of the host material of a wire. This translates into $e^2/\hbar\kappa v_F\sim1$ and  $E_F\sim10$ meV so that for the temperature range $T\sim0.25\div1$ K the thermal relaxation length is of the order $\sim2.5\div10$ $\mu$m.
\end{thebibliography}
\end{document}